\documentclass{article}
\usepackage{spconf,amsmath,graphicx}
\usepackage{multirow}
\usepackage[flushleft]{threeparttable} % For tablenote

\title{Multiclass Arrhythmia Classification using Smartwatch Photoplethysmography Signals Collected in Real-life Settings}
\name{\parbox{\linewidth}{\centering
        Dong Han\textsuperscript{1} \qquad
        Jihye Moon\textsuperscript{1} \qquad
        Luís Roberto Mercado Díaz\textsuperscript{1} \qquad
        Darren Chen\textsuperscript{1} \qquad
        Devan Williams\textsuperscript{2} \qquad
        Eric Y. Ding\textsuperscript{3} \qquad
        Khanh-Van Tran\textsuperscript{4} \qquad
        David D. McManus\textsuperscript{4} \qquad
        Ki H. Chon\textsuperscript{1}\sthanks{This work was supported by NIH under Grant R01 HL137734.}
        }
    }
\address{\parbox{\linewidth}{\centering
        \textsuperscript{1}University of Connecticut \qquad
        \textsuperscript{2}Morehouse School of Medicine \qquad
        \textsuperscript{3}The Warren Alpert Medical School of Brown University \qquad
        \textsuperscript{4}University of Massachusetts Chan Medical School
            }
        }
\begin{document}
%\ninept
%
\maketitle
% % --- Luis change 03/ start ---
% \vspace{-1.5em} % Reduces the space between title and abstract
% % --- Luis change 03/ end ---
%
\begin{abstract}

Most deep learning models of multiclass arrhythmia classification are tested on fingertip photoplethysmographic (PPG) data, which has higher signal-to-noise ratios compared to smartwatch-derived PPG, and the best reported sensitivity value for premature atrial/ventricular contraction (PAC/PVC) detection is only 75\%. To improve upon PAC/PVC detection sensitivity while maintaining high AF detection, we use multi-modal data which incorporates 1D PPG, accelerometers, and heart rate data as the inputs to a computationally efficient 1D bi-directional Gated Recurrent Unit (1D-Bi-GRU) model to detect three arrhythmia classes. We used motion-artifact prone smartwatch PPG data from the NIH-funded Pulsewatch clinical trial. Our multimodal model tested on 72 subjects achieved an unprecedented 83\% sensitivity for PAC/PVC detection while maintaining a high accuracy of 97.31\% for AF detection. These results outperformed the best state-of-the-art model by 20.81\% for PAC/PVC and 2.55\% for AF detection even while our model was computationally more efficient (14 times lighter and 2.7 faster).

\end{abstract}
\section{Introduction}
\label{sec:intro}

% These guidelines include complete descriptions of the fonts, spacing, and
% related information for producing your proceedings manuscripts. Please follow
% them and if you have any questions, direct them to Conference Management
% Services, Inc.: Phone +1-979-846-6800 or email
% to \\\texttt{papers@2024.ieeeicassp.org}.
Atrial fibrillation (AF) is the most prevalent malignant cardiac dysrhythmia. The prevalence of AF in the U.S. is expected to rise to 12.1 million (projections to 3.4\% of the total population) in 2030 as the population ages \cite{viraniHeartDiseaseStroke2021}. Long-term monitoring for AF is effective for incident AF detection, as most cases of early stages of AF are brief and intermittent \cite{flakerAsymptomaticAtrialFibrillation2005}. However, existing conventional gold standard methods for continuous AF monitoring via ECG suffer from poor patient acceptability and low long-term adherence, largely due to their reliance on gel electrodes and multiple leads. An alternate solution for continuous monitoring is via a smartwatch with photoplethysmography (PPG), which provides a convenient solution for continuous monitoring of AF \cite{chongArrhythmiaDiscriminationUsing2015}. Recent works using smartwatches have shown accurate AF detection \cite{lubitzDetectionAtrialFibrillation2022, perezLargeScaleAssessmentSmartwatch2019, avramValidationAlgorithmContinuous2021}, but most algorithms are not able to discriminate premature atrial and ventricular contractions (PAC/PVC) \cite{dingPhotoplethysmographyBasedAtrial2024}. It is important to distinguish PAC and PVC, as frequent occurrences of these rhythms can mimic AF dynamics, thereby reducing accuracy of AF detection \cite{basharAtrialFibrillationDetection2019c}.

While it is relatively easy to detect PAC and PVC in ECG signals \cite{basharAtrialFibrillationDetection2020}, it is rather difficult to detect these rhythms in PPG due to inexact signature waveforms representing these arrhythmias \cite{dingPhotoplethysmographyBasedAtrial2024}. Another challenge with PPG for arrhythmia detection is that motion noise artifacts are a significant issue in smartwatch data, as they are known to distort the PPG waveforms and mimic irregular dynamics seen in AF \cite{basharAtrialFibrillationDetection2019c}. To combat these two issues, more comprehensive smartwatch databases are needed to account for diverse sets of motion artifacts and to have a sufficient number of PAC and PVC cases to train deep learning methods. However, long duration recordings of smartwatch PPG data require time-consuming adjudication of AF and PAC/PVC rhythms, using simultaneous recordings of ECG signals as the gold standard.

A further complication of using a smartwatch for arrhythmia detection is that most prior work relied on data collected in a controlled environment (typically with minimum or unrealistic motion artifacts) or PPG data from a fingertip, which exhibits higher signal-to-noise ratios (SNR) and less motion noise than does smartwatch PPG data \cite{rajalaComparisonPhotoplethysmogramMeasured2018}. One of the first works on PAC/PVC detection from a fingertip PPG is by Poh et al. \cite{pohDiagnosticAssessmentDeep2018}, which reported 72.2\% sensitivity and 85.0\% positive predictive value (PPV) using a 1D-DenseNets model. However, testing results were based on the same training subjects instead of using independent subjects from the training datasets. A work by Liu et al. \cite{liuMulticlassArrhythmiaDetection2022} reported subject-independent testing with 75.4\% sensitivity and 82.7\% PPV for PAC/PVC detection using a 1D VGG-16 model on motion-free fingertip PPG data. Since these fingertip PPG data were recorded in clinics with minimal motion artifacts, it is not likely that the same results can be achieved when real-life smartwatch PPG data are used. 

To address the current limitations of accurate PAC/PVC detection from PPG, we recently conducted an NIH-funded study called “Pulsewatch” \cite{dingAccuracyUsabilityAdherence2023} which continuously recorded PPG data from older stroke survivors in their homes via smartwatches with simultaneous reference ECG for 14 days. In this work, we aim to improve the sensitivity of PAC/PVC detection, while maintaining a high accuracy on AF detection using multi-modal data, including PPG, heart rates (HRs), and accelerometer (ACC), with a computationally efficient deep learning model. We also aim to demonstrate the generalizability of our deep learning model by using many independent testing subjects and many PPG segments.

\section{Dataset Description}
\label{sec:dataset}
\subsection{Pulsewatch dataset}
\label{subsec:pulsewatchdataset}

We recently conducted an NIH-funded clinical trial called Pulsewatch\footnote{The data used in this work will be available on https://www.synapse.org/Synapse:syn23565056/.} to determine the feasibility of detecting AF using smartwatches in real-life conditions \cite{dingAccuracyUsabilityAdherence2023}. Participants (n=106) in the Pulsewatch clinical trial continuously wore the smartwatch system (which also included a smartphone for data collection) with a reference ECG chest patch for 14 days in their everyday lives. Demographic and medical history information of the recruited participants (aged $\geq$50 with a history of ischemic stroke) can be found in \cite{dingAccuracyUsabilityAdherence2023}. Formal ethical approval for this study was obtained from the University of Massachusetts Medical School Institutional Review Board (approval numbers H00016067 and H00009953). Written informed consent was collected from all patient participants. The smartwatch system (Samsung Gear S3 or Galaxy Watch 3, Samsung, San Jose, CA, USA) recorded single-channel PPG and accelerometer signals at 50 Hz \cite{hanSmartwatchSystemContinuous2023}. Arrhythmias were adjudicated in each 30-second segment by three experts \cite{tranFalseAtrialFibrillation2023a} using the aligned single-channel ECG (Cardea SOLO, Cardiac Insight Inc., Bellevue, WA, USA) that was sampled at 250 Hz.

We used our previously developed automated motion and noise detection algorithm \cite{mohagheghianOptimizedSignalQuality2022a} on the Pulsewatch data, and only those segments that were determined to be relatively clean PPG segments ($\leq$5 seconds of motion noise) were used for subsequent multiclass arrhythmia classification. This tolerance of $\leq$5 sec of motion noise was used primarily to increase the number of usable PPG data segments for arrhythmia detection, as we found in our previous study that this did not cause many false AF alerts \cite{basharAtrialFibrillationDetection2019c, mohagheghianOptimizedSignalQuality2022a}.
\vspace{-1em}

\subsection{Training and testing datasets}
\label{subsec:traintestsplit}
The number of segments with AF and PAC/PVC differ among subjects, with some subjects having many instances of these rhythms while others had few to none. Hence, how to determine which subjects to use for training and which for subject-independent testing became a challenging issue for addressing the generalizability of the proposed method. Therefore, given the imbalanced datasets, we used a two-fold cross-validation and divided the entire Pulsewatch dataset into two equal halves based on all AF and PAC/PVC subjects in the dataset. Table \ref{table_subject} shows that each fold had nearly the same number of subjects and segments for all arrhythmia classes. 

% \usepackage{multirow}
\begin{table}[]
\caption{Subject and Data Segment Information for Two-fold Cross Validation for the Pulsewatch Dataset}
\label{table_subject}
\resizebox{\columnwidth}{!}{
\begin{tabular}{llllll}
\hline
                          & Fold name & NSR    & AF     & PAC/PVC & Total   \\ \hline
\multirow{3}{*}{Subjects} & Fold-1    & 18     & 17     & 19      & 36      \\
                          & Fold-2    & 18     & 18     & 19      & 36      \\
                          & Total     & 36     & 35     & 38      & 72      \\ \hline
\multirow{3}{*}{Segments} & Fold-1    & 39,356 & 12,265 & 6,697   & 58,318  \\
                          & Fold-2    & 39,363 & 12,290 & 6,342   & 57,995  \\
                          & Total     & 78,719 & 24,555 & 13,039  & 116,313 \\ \hline
\end{tabular}
}
\begin{tablenotes}
      \small
      \item NSR: normal sinus rhythm; AF: atrial fibrillation; PAC/PVC: premature atrial contraction/premature ventricular contraction.
\end{tablenotes}
\vspace{-2em}
\end{table}

For each fold, we performed subject-based stratified random sampling to divide the data into 80\% training, 10\% validation, and 10\% subject-dependent testing. For the subject-independent testing results, the data from the opposite fold was used to test the fine-tuned model from one of the folds, as subjects were mutually exclusive in each fold.
% % --- Luis change 04/ start ---
\vspace{-0.5em} % Reduces the space between title and abstract
% % --- Luis change 04/ end ---

\section{Methods}
\label{sec:methods}
% Below is an example of how to insert images. Delete the ``\vspace'' line,
% uncomment the preceding line ``\centerline...'' and replace ``imageX.ps''
% with a suitable PostScript file name.
% -------------------------------------------------------------------------
\begin{figure*}[htb]
  \centering
  \centerline{\includegraphics[width=\textwidth]{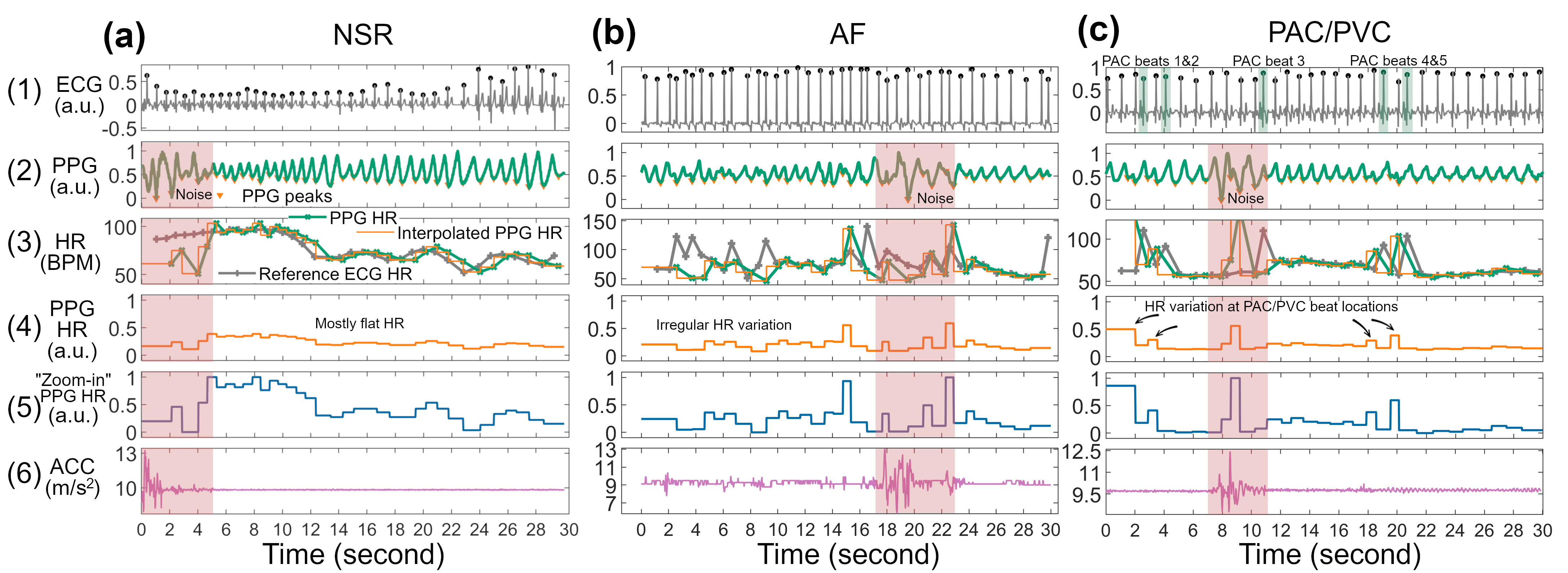}}
\caption{Example 30-sec segments from (a) NSR, (b) AF, and (c) PAC/PVC with less than 5 sec of motion noise artifact recorded from the Pulsewatch system. Rows: (1) reference ECG, (2) filtered PPG, (3) heart rates (HR) calculated from the peaks of ECG and PPG, and the interpolated PPG HR that was used for our proposed model, (4) normalized PPG HR with minimum and maximum values from 30 to 220 BPM range, (5) “zoom-in” PPG HR for regular heart rate to better accentuate dynamic ranges, and (6) accelerometer (ACC) signals. Signals from rows (2), (4), (5), and (6) were used in our best proposed model.}
\vspace{-0.9em} % Reduces the space between caption and following text.
\label{fig:signal}
\end{figure*}

While most prior works used complicated deep learning models for multiclass arrhythmia classification (Poh et al.’s 1D-DenseNets \cite{pohDiagnosticAssessmentDeep2018}, Liu et al.’s 1D VGG-16 \cite{liuMulticlassArrhythmiaDetection2022}, and Chen et al.’s 2D-DenseNets \cite{chenSmartwatchPhotoplethysmogramBasedAtrial2024}), a recent work \cite{sabourGatedRecurrentUnitBased2022} has shown that a simpler model structure such as the 1D bi-directional Gated Recurrent Unit (1D-bi-GRU) can provide accurate detection of motion artifacts in PPG data. Furthermore, the same model structure trained on 1D PPG signals worked moderately better than did a 2D time-frequency spectrogram (TFS) \cite{sabourGatedRecurrentUnitBased2022}. Hence, we used a 1D time series as the input and implemented the 1D-Bi-GRU model as described in \cite{sabourGatedRecurrentUnitBased2022}. This model is particularly well suited for capturing temporal dependencies in PPG signals, which are critical for distinguishing subtle dynamics present in cardiac arrhythmias.

We also added HR as an input because our prior work has shown that cardiac arrhythmias can be accurately discriminated using HRs \cite{hanPrematureAtrialVentricular2020a}. Moreover, ACC signals were used to train the network as motion artifacts could occur so that the network learned to ignore those contaminated PPG segments.
\vspace{-1.5em}

\subsection{Signal Processing of the Time-Series Data}
\label{subsec:signalproc}
\subsubsection{1D Time Series Data Preparation}
\label{subsubsec:1Dtimeseriesprep}
The left, middle, and right top rows of Fig.\ref{fig:signal} show representative ECG signals for normal sinus rhythm (NSR), AF, and PAC/PVC, respectively. Row (2) of Fig.\ref{fig:signal} shows the corresponding and simultaneously measured PPG, filtered with a 6$^{th}$-order Butterworth bandpass IIR filter (0.5 to 20 Hz) \cite{mohagheghianNoiseReductionPhotoplethysmography2024}. Each filtered PPG was normalized to [0, 1] based on each segment’s minimum and maximum values. The third row shows HRs obtained via ECG and via the corresponding PPG along with interpolated PPG HR (shown in orange lines), which is used to account for the abrupt HR changes. The fourth and fifth rows show normalized PPG heart rates and magnified PPG heart rates, respectively, where the fourth row was normalized within a [30, 220] BPM range to represent those with rapid ventricular response (RVR) (e.g., heart rates $>$100 BPM) \cite{hanRealTimePPGPeak2022a}. The fifth row represents each segment’s minimum and maximum HR values so that non-RVR rhythms can be represented with better dynamic ranges. The tri-axial accelerometers’ (ACC) magnitudes in the 0 to 20 $m/s^2$ range (daily activity range) are shown in row 6 of Fig.\ref{fig:signal}. 
\vspace{-1em}

\subsubsection{Extraction of PPG heart rates (HR)}
\label{subsec:extractHR}
The HR of each PPG segment was automatically calculated using the WEPD PPG peak detection algorithm as this approach has been shown to be one of the most accurate and can account for various arrhythmias \cite{hanRealTimePPGPeak2022a}. 
\vspace{-1em}

\subsection{Machine Learning Model Design: 1D-Bi-GRU Model}
\label{subsec:mlmodel}
\begin{figure}[htb]
  \centering
  \centerline{\includegraphics[width=0.85\columnwidth]{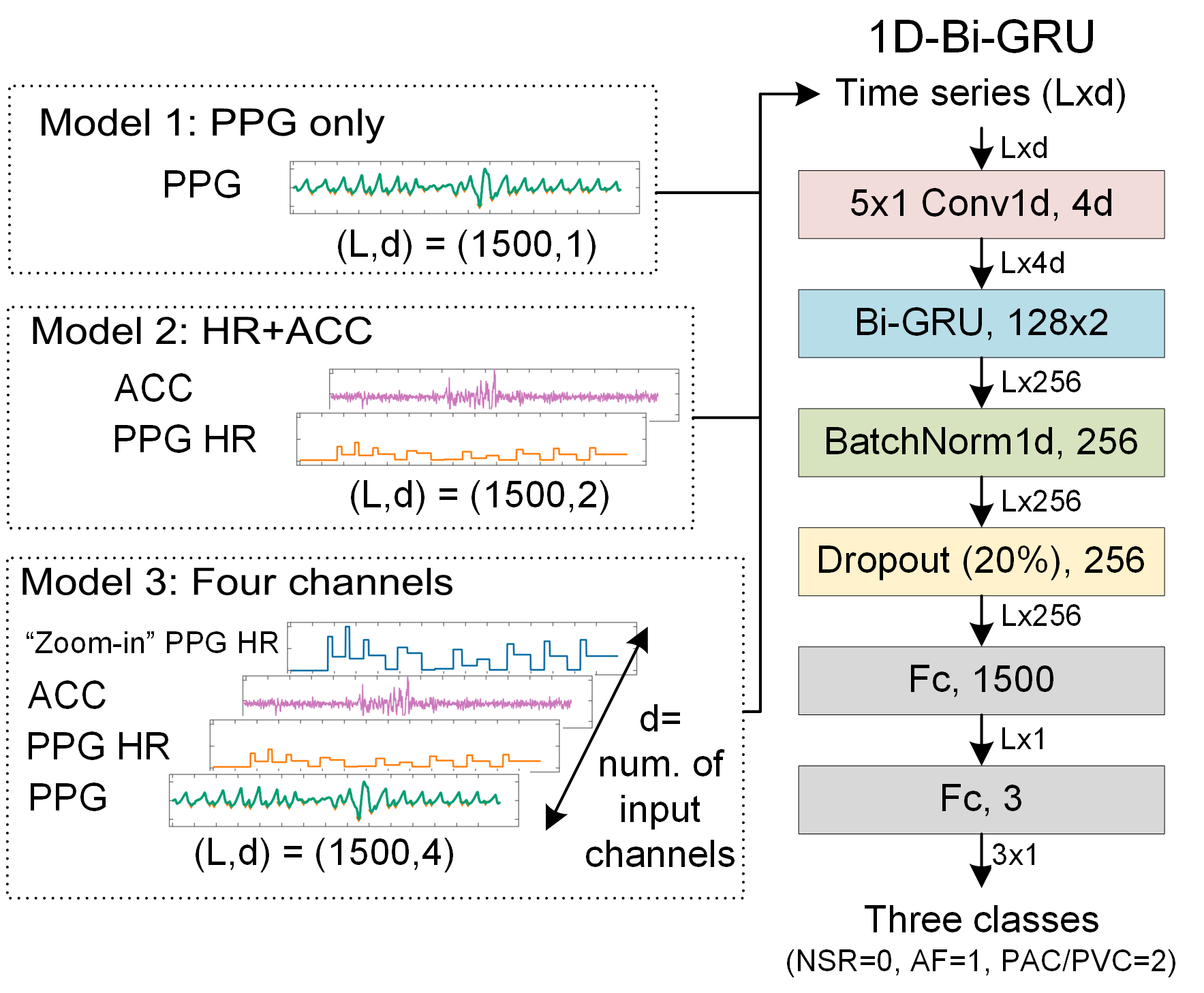}}
\caption{The architecture of our proposed 1D bi-directional Gated Recurrent Unit (1D-Bi-GRU) model.}
\vspace{-1.5em} % Reduces the space between caption and following text.
\label{fig:model}
\end{figure}

As described in \cite{sabourGatedRecurrentUnitBased2022} and shown in Fig.\ref{fig:model}, our input time series has a dimension of ($L$, $d$) ($L$=1,500 samples in our case, while $d$ is the number of input channels). The first layer is a 1D convolutional neural network (CNN) to embed the input time series with $4d$ filters with a kernel size of 5, a stride size of 1, and a padding size of 2 to ensure the output dimension is ($L$, $4d$). The second layer is a bi-GRU layer, which combines the outputs of two GRU networks (with 128 units each) that process the input-embedded information in the opposite directions, allowing for each sample to consider both preceding and proceeding samples. A batch normalization is then applied, followed by a dropout of 20\% to avoid overfitting. Lastly, a dense layer combines the output of the previous layers ($L$, $256$) into a dimension of ($L$, $3$) for predicting three classes (0=NSR, 1=AF, and 2=PAC/PVC). 

% Below is an example of how to insert images. Delete the ``\vspace'' line,
% uncomment the preceding line ``\centerline...'' and replace ``imageX.ps''
% with a suitable PostScript file name.
% -------------------------------------------------------------------------

\subsection{Machine Learning Model Training Process}
\label{subsec:mltrain}
As shown in Table \ref{table_subject}, the number of NSR segments is 3 and 5 times more than AF and PAC/PVC segments, respectively. To prevent over-fitting, up-sampling of the minority classes was implemented in the training and validation sets to ensure unbiased performance in the testing data.

A batch size of 32 was used, as it showed a faster and more stable training validation process compared to a batch size of 512. The cross-entropy loss function was used for three-class classification. The Adam optimizer was used with the same parameters described in \cite{sabourGatedRecurrentUnitBased2022}. We trained the models\footnote{The code will be available on https://github.com/Cassey2016.} with a maximum of 200 epochs and selected the best model using the minimum validation loss. Early stop was used if the validation loss did not improve in 40 consecutive epochs.
\vspace{-1em}

\subsection{Evaluation Metrics}
\label{subsec:evalmetrics}
For the performance evaluation of the proposed and other compared methods, we calculated five key metrics: sensitivity (sens.), specificity (spec.), precision (prec.), negative predictive value (NPV), and accuracy (accu.) in keeping with other publications \cite{pohDiagnosticAssessmentDeep2018, liuMulticlassArrhythmiaDetection2022, hanPrematureAtrialVentricular2020a}. We performed subject-independent testing as detailed in section \ref{subsec:traintestsplit}.
\vspace{-0.5em}

\section{Results and Discussion}
\label{sec:resultsdiscussion}

\subsection{Effectiveness of using HR as an Input for Multiclass Arrhythmia Classification}
\label{subsec:HRasinput}

Table \ref{table_metrics} shows the comparison of two other methods versus our proposed 1D-Bi-GRU model with different combinations of inputs for the multiclass classification results. As shown, the best performance metrics were with our proposed model with four time series (1D PPG, HR, ACC, and “zoomed-in” HR) as the inputs. To date, the best reported sensitivity for PAC/PVC detection is 75.4\% by Liu et al. \cite{liuMulticlassArrhythmiaDetection2022}, and we implemented Liu et al. and Chen et al.’s models, retrained them using the same data, and reported the results in the first and second rows of Table \ref{table_metrics}, for proper comparison. The best sensitivity for PAC/PVC after retraining Liu et al.’s model \cite{liuMulticlassArrhythmiaDetection2022} is only 63\%, as shown in the first row of Table \ref{table_metrics}. However, our approach resulted in 84\% sensitivity for PAC/PVC detection. Note that HR information is especially needed for higher performance metrics; when only PPG was used as the input signal, the PAC/PVC detection sensitivity was only 67\%. It increased to $>$81\% when HR data are added, as shown in the bottom two rows of Table \ref{table_metrics}.

% \usepackage{multirow}
\begin{table}[]
\caption{Subject Independent Testing Results on the Pulsewatch Two-fold Cross-Validation Data}
\label{table_metrics}
\resizebox{\columnwidth}{!}{
\begin{tabular}{llllllll}
\hline
                                                                                 & Methods                                                                                & Rhythm  & Sens. & Spec. & Prec. & NPV   & Acc.  \\ \hline
\multirow{6}{*}{\rotatebox[origin=c]{90}{\begin{tabular}[c]{@{}l@{}}Previous \\ (retrained)\end{tabular}}} & \multirow{3}{*}{\begin{tabular}[c]{@{}l@{}}Liu et al. \cite{liuMulticlassArrhythmiaDetection2022}\\ (2022)\end{tabular}}  & NSR     & \bf{96.03} & 91.25 & 95.83 & \bf{91.65} & \bf{94.48} \\
                                                                                 &                                                                                        & AF      & 90.21 & 95.98 & 85.73 & 97.34 & 94.76 \\
                                                                                 &                                                                                        & PAC/PVC & 63.34 & \bf{96.78} & \bf{71.27} & 95.44 & \bf{93.03} \\ \cline{2-8} 
                                                                                 & \multirow{3}{*}{\begin{tabular}[c]{@{}l@{}}Chen et al. \cite{chenSmartwatchPhotoplethysmogramBasedAtrial2024}\\ (2024)\end{tabular}} & NSR     & 92.88 & 91.33 & 95.73 & 85.97 & 92.38 \\
                                                                                 &                                                                                        & AF      & 89.21 & 96.34 & 86.70 & 97.09 & 94.83 \\
                                                                                 &                                                                                        & PAC/PVC & 59.43 & 93.30 & 52.82 & 94.80 & 89.50 \\ \hline
\multirow{9}{*}{\rotatebox[origin=c]{90}{Ours}}                                                            & \multirow{3}{*}{1D PPG only}                                                           & NSR     & 90.20 & 95.66 & 97.75 & 82.33 & 91.96 \\
                                                                                 &                                                                                        & AF      & 89.65 & 93.93 & 79.81 & 97.14 & 93.03 \\
                                                                                 &                                                                                        & PAC/PVC & 67.47 & 92.93 & 54.64 & 95.77 & 90.08 \\ \cline{2-8} 
                                                                                 & \multirow{3}{*}{HR and ACC}                                                            & NSR     & 86.95 & 95.15 & 97.40 & 77.68 & 89.60 \\
                                                                                 &                                                                                        & AF      & 97.03 & 96.65 & 88.58 & 99.18 & 96.73 \\
                                                                                 &                                                                                        & PAC/PVC & 80.90 & 91.67 & 55.09 & 97.44 & 90.46 \\ \cline{2-8} 
                                                                                 & \multirow{3}{*}{\begin{tabular}[c]{@{}l@{}}Four channels\\ (best)\end{tabular}}        & NSR     & 88.25 & \bf{96.18} & \bf{97.98} & 79.62 & 90.81 \\
                                                                                 &                                                                                        & AF      & \bf{97.52} & \bf{97.25} & \bf{90.47} & \bf{99.32} & \bf{97.31} \\
                                                                                 &                                                                                        & PAC/PVC & \bf{83.52} & 92.20 & 57.48 & \bf{97.79} & 91.23 \\ \hline
\end{tabular}
}
\vspace{-1.5em}
\end{table}

It should be noted that the approach by Liu et al. \cite{liuMulticlassArrhythmiaDetection2022} used fingertip PPG data which has a higher SNR, but the method by Chen et al. \cite{chenSmartwatchPhotoplethysmogramBasedAtrial2024} used a different subset of the Pulsewatch dataset with 2D spectrum of PPG as the input which had a much higher computational cost. The reported sensitivity to PAC/PVC of Chen et al.’s model was only 66.40\% \cite{chenSmartwatchPhotoplethysmogramBasedAtrial2024} and now 59.43\% after retraining with the current data, which is lower than Liu et al.’s model and our model that uses only 1D PPG waveform as the input. Despite the fact that our method used smartwatch PPG, the performance metrics for AF and PAC/PVC detection are much higher than those reported by Liu et al. \cite{liuMulticlassArrhythmiaDetection2022}. More importantly, our testing results were derived from a large number of independent subjects. Hence, the results with our proposed approach highlight the potential better generalizability of the model, as we used a far greater number of segments, PPG signals with lower SNR than those of fingertip PPG, and the model was tested with independent subjects from the training dataset. It should be noted that our approach showed lower sensitivity (88.25\%) for NSR detection compared to Liu’s model, with 96.03\% sensitivity. We believe this is because our model incorrectly selected some of the NSR as PAC/PVC and we suspect this might be attributed to motion artifacts causing incorrect HR estimation. Note that clinically this is not a significant concern since NSR is not a malignant rhythm.
\vspace{-0.5em} % Reduces the space between paragraph and the next subtitle.

\subsection{Efficiency in Computational Cost using HR as an Input}
\label{subsec:compucost}
Our model is computationally efficient, potentially allowing for real-time classification of cardiac arrhythmias on wearable devices. Table \ref{table_compucost} shows that our best proposed model has only 120,224 parameters, which is only 1/14$^{th}$ of the number of parameters used in Liu’s model \cite{liuMulticlassArrhythmiaDetection2022}. Our best proposed model only uses 0.89 billion floating-point operations per second (GFLOPs where G=giga/billion), which is 2.7 times faster than Liu’s model \cite{liuMulticlassArrhythmiaDetection2022}.
\vspace{-0.5em}

\begin{table}[]
\caption{Computational Cost for Different Methods}
\label{table_compucost}
\centering
\resizebox{0.8\columnwidth}{!}{
\begin{tabular}{lll}
\hline
Methods              & Num. of parameters & GFLOPs \\ \hline
Liu et al. \cite{liuMulticlassArrhythmiaDetection2022}  & 1,626,403          & 2.36   \\
Chen et al. \cite{chenSmartwatchPhotoplethysmogramBasedAtrial2024} & 221,303            & 19.33  \\ \hline
Ours (1D PPG only)   & 110,696            & 0.75   \\
Ours (HR and ACC)    & 113,832            & 0.77   \\
Ours (four channels) & 120,224            & 0.89   \\ \hline
\end{tabular}
}
\vspace{-1.5em}
\end{table}

\section{CONCLUSION}
\label{sec:conclusion}

In this work, we proposed a computationally efficient 1D-Bi-GRU model to classify NSR, AF, and PAC/PVC using smartwatch PPG data collected in real-life settings. Our model showed accurate AF classification as well as the highest PAC/PVC sensitivity value ever reported in the literature, besting even those studies which used higher SNR-based PPG data collection modalities (e.g., fingertip PPG). Our better performance metrics were mainly due to the addition of HR as an input to the 1D-Bi-GRU model.

%\vfill
%\pagebreak

\vfill\pagebreak

% \section{REFERENCES}
\label{sec:refs}

% % --- Luis change 05/ start ---
% % Before including bibliography
% \setstretch{0.8} % Reduces the line spacing in references to 1 (single spacing)
% \setlength{\bibsep}{0pt plus 0.3ex}  % Reduces spacing between references

% % References
% % --- Luis change 05/ end ---
\bibliographystyle{IEEEbib}
\bibliography{myref}

\end{document}